\documentclass
[aps,prl,twocolumn,showpacs,floats,floatfix,amsmath]{revtex4}
\usepackage{graphicx}
\usepackage{amsmath}
\usepackage{amsfonts}
\usepackage{amssymb}%
\usepackage{epsf}
\usepackage{rotate}

\newcommand{\be}{\begin{equation}}
\newcommand{\ee}{  \end{equation}}
\newcommand{\ba}{\begin{eqnarray}}
\newcommand{\ea}{  \end{eqnarray}}
\newcommand{\ket}[1]{\left|#1\right>}
\newcommand{\bra}[1]{\left< #1 \right|}

\begin{document}

\title{Loschmidt echo in the Bose-Hubbard model: 
turning back time in an optical lattice}

\author{Fernando M. Cucchietti}
\affiliation{%
Theoretical Division, Los Alamos National Laboratory, Los Alamos, NM 87545, USA
}%

\date{September 20, 2006}%
\begin{abstract}
I show how to perform a Loschmidt echo (time reversal) 
in the Bose-Hubbard model implemented with cold bosonic atoms
in an optical lattice. The echo is obtained by applying a linear phase imprint
on the lattice and a change in magnetic field to tune the boson-boson 
scattering length through a Feshbach resonance.
I discuss how the echo can measure the fidelity of the quantum simulation,
the intensity of an external potential (e.g. gravity), or the critical point 
of the superfluid-insulator quantum phase transition.
\end{abstract} 

\pacs{03.75.Lm,39.25.+k,32.80.Qk}

\maketitle 

More than two decades ago \cite{Feynman}, Feynman envisioned 
quantum physics simulations being performed by
controllable quantum systems  as
no traditional computer could. 
For large complex systems, it is not always practical to estimate
the accuracy of the simulation, given by the fidelity $f(t)$ of the actual
experimental evolution with respect to the desired one.
Yet, in some cases $f(t)$ can be measured directly.
Let ${\cal H}_{sim}$ be the Hamiltonian to be simulated and 
${\cal H}={\cal H}_{sim}+V$ the laboratory Hamiltonian, 
where $V$ contains unknown terms present in the real experiment. 
To measure $f(t)$, one has to
first evolve a given initial state $\ket{\psi_0}$ with ${\cal H}$, perform an operation
that {\em changes the sign}
of ${\cal H}_{sim}$, evolve for another time $t$ with the new Hamiltonian 
 ${\cal H}^{'} =-{\cal H}_{sim}+V$ \cite{perturbations}, and then measure the 
 probability of finding the system in the initial state,
 \be
 f(t)=|\bra{\psi_0} e^{-i (-{\cal H}_{sim}+V)t} e^{-i({\cal H}_{sim}+V))t} \ket{\psi_0}|^2.
 \label{loschmidt}
 \ee
Clearly, $\left.f(t)\right|_{V=0}=1$. In general, 
$f(t)$ -- also known as a Loschmidt echo (LE) -- depends on $V$ \cite{Peres}. 
Therefore, to measure $f(t)$ one only needs the ability to 
change the sign of the Hamiltonian.
Notice that, in quantum mechanics,
this operation is equivalent to reversing time.
Thus, the LE is the probability of return to the initial state
by a forward-backwards evolution in presence of a perturbation.

Recently, the Bose-Hubbard model (BHM) was simulated using
cold bosonic atoms loaded in an optical lattice \cite{CiracBHM,Greiner}.
Many models can be simulated with this system \cite{Lewenstein}. 
However, the BHM has a strong appeal because of its simplicity
and because it presents a superfluid-insulator transition --
a paradigm of quantum phase transitions \cite{Sachdev}. 
In this letter I propose an experimental procedure to change the sign of the 
BHM Hamiltonian and perform a Loschmidt echo.
The time reversal operation consists of a phase imprinting in the lattice and
a sign change of the boson-boson scattering length by varying the magnetic
field near a Feshbach resonance -- both techniques have been demonstrated 
\cite{Greiner,PhillipsPhase,Feshbach}.
Also important, when the fidelity of the simulation is high, the LE can be a sensing tool: 
I will show how it can be used to measure e.g. the intensity of external potentials, 
or the critical point of the BHM quantum phase transition.

A simple example of a LE is the Hahn or spin echo 
\cite{Hahn}. In a typical nuclear magnetic 
resonance (NMR) experiment, the polarization signal decays rapidly 
due to the inhomogeneity of the local magnetic field -- spins 
precess at different Larmor frequencies. 
However, a $\pi$ rotation of the 
spins changes the effective sign of the spin-magnetic field interaction, 
refocusing the polarization as if time were reversed.  
The spin echo decay is used to measure the relaxation time $T_2$, given by 
other terms in the full Hamiltonian (e.g. spin-spin interactions) that
are not reversed by the $\pi$ rotation.
More complex LEs have been performed e.g. in solid state NMR \cite{Horacio}
(including many-body interaction terms), 
and in trapped cold atomic systems \cite{Israel}.
In general, the LE is a measure of the stability of quantum evolution
with respect to perturbations \cite{Peres}. For instance, in a classically
chaotic system there is a regime where the the LE decays with 
the Lyapunov exponent of the classical Hamiltonian \cite{Jalabert}. 
Also, in open systems the decay rate of the LE equals the  
decoherence rate \cite{CucchiettiPRL}.
Recently, it was shown that the decay of the LE can signal
a quantum phase transition in the environment 
\cite{Quan,CucchiettiQPT,Rossini,Zanardi}.
For pure states, the LE is equal to 
the fidelity of a quantum computation \cite{Nielsen}.

The BHM Hamiltonian is \cite{Fisher}
\be
{\cal H}_{BH}=-J \sum_{<i,j>} a_i^\dagger a_j + U \sum_i n_i (n_i - 1),
\label{BH}
\ee
where $a_i$ are bosonic annihilation operators in the $i^{th}$ site of a discrete lattice,
$J$ is the hopping amplitude between neighboring sites,
and $U$ is the interaction energy of bosons in the same site. 
The BHM undergoes a quantum phase transition when $J/U$ is varied \cite{Sachdev}.
For $J \gg U$, the dominating hopping term 
favors delocalization of particles: the ground state is a superfluid. 
In the opposite regime, $J \ll U$, the interaction energy is too strong and
number fluctuations in a site are costly:
the ground state (for integer density) is a gapped Mott insulator.
The BHM is implemented using cold atoms in a periodic oscillatory 
potential created with a standing wave laser light (the optical lattice) \cite{CiracBHM}.
Up to good approximation,
the atoms occupy only the ground state of each well of the lattice. 
The overlap between ground states of adjacent sites gives  the tunneling amplitude $J$.
Longer neighbor tunneling is suppressed exponentially. 
A contact interaction potential with a range much smaller than the size of the wells is assumed. 
In the right limit, this translates in the interaction term of Eq. (\ref{BH}),
with $U$ directly proportional to the $s$-wave boson-boson scattering length $a_s$.
In all implementations of the BHM in optical lattices so far there is
an additional magnetic trapping potential \cite{Greiner}.
To perform a LE it would also be necessary to change the sign of this 
potential (the system remains stable for a time equal to the forward in time evolution). 
However, an inhomogeneous magnetic field implies
a position dependent interaction strength $U$, departing from the simplest BHM. 
For simplicity, I will only consider the homogeneous BHM, which will be realized
experimentally in the near future \cite{Raizen,PrivCom}. In addition, I will only consider 
systems with open boundary conditions. The experiment proposed here 
could be applied to other implementations of the  BHM, e.g. with trapped 
ions \cite{CiracOL}, or even in the Hubbard model with fermions.

\begin{figure}[b]
\centering \leavevmode
\epsfxsize 3.2in
\epsfbox{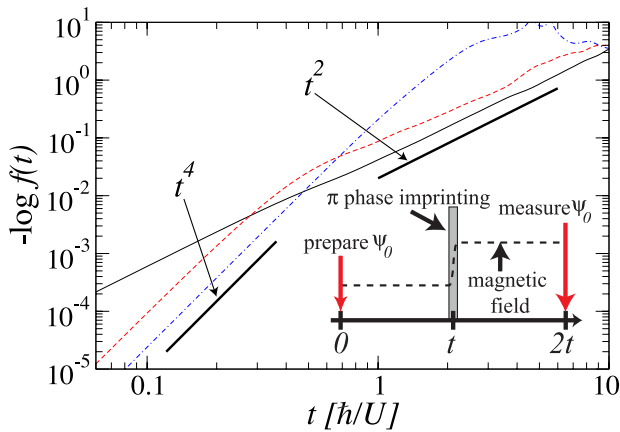}
\caption{Loschmidt echo for a perturbation in  $J$ (solid line) with $\delta J/J=0.05$, 
in $U$ (dashed line) with $\delta U/U=0.2$, and for a gravity potential (dash-dotted line)
$mdg/U=0.1$. For comparison, the thick lines show the
$t^4$ and $t^2$ dependence. All the curves are for a 7 site
system with unit density, and $J/U=1$. Similar behavior is observed 
for a wide variety of initial states.
In the inset, schematic of the time reversal sequence. 
The initially prepared state $\psi_0$ is evolved with the optical lattice
Hamiltonian for a time  $t$, where the sign of the BHM Hamiltonian is changed by 
a phase imprinting gate and a change in the external magnetic field.
After another time $t$, the probability of finding the system in $\psi_0$ is 
measured, Eq. (\ref{loschmidt}). If $f(t)=1$, the time reversal was perfect. 
}
\label{Figure1}
\end{figure}

The change of sign of ${\cal H}_{BH}$ is done in separate steps for $J$ and $U$. 
The change of $U$ is achieved using a marvelous experimental handle of cold atomic systems: 
the effective atom-atom scattering length can be {\em tuned} by varying an external magnetic 
field near a Feshbach resonance \cite{Feshbach}. 
The external field shifts the energy of the free atoms with respect to a bound molecular state that
changes drastically the scattering process.
In the simplest cases, the scattering length as a function of the magnetic field $B$ 
is $a_s(B)=a_{bg}\left(1-\frac{\Delta B}{B-B_0}\right)$, 
where $a_{bg}$ is the scattering length of atoms in absence of the quasi-bound state, 
$B_0$ is the resonance position (related to the energy of the bound state)
and $\Delta B$ is the width of the resonance.
Therefore, to flip the sign of $U$
one needs to rapidly change the external magnetic field so that the
Feshbach resonance is crossed and $a_s \rightarrow -a_s$. 

The hopping amplitude $J$ is proportional to the overlap between the ground 
states of neighboring sites \cite{CiracBHM}. Thus, in principle, it cannot change sign. 
However, like the $\pi$ pulse in the spin echo example above \cite{Hahn},
one can apply an operation that changes the effective sign of the
Hamiltonian acting on the wave function.  
The spectrum of the hopping term of the BHM (i.e for $U=0$) in each dimension is 
$E(k)=-2J \cos(k d)$, where $d$ is the lattice site spacing,
$k=n \pi /d(N+1)$ with $n=1..N$, and $N$ is the number of sites in the lattice.
Clearly, the sign of $E(k)$ is changed
by boosting all $k$ states by momentum $\pi/d$, i.e. $E(k+\pi/d)=-E(k)$.
The $p$ momentum translation operator is diagonal in real space,
$\hat P_p=\exp(-i p d \sum_j j \hat n_j)$: this is
equivalent to the evolution operator under a linear potential 
$F \sum_j j \hat n_j$ for a time $\tau=pd/F$.
Thus, the sign of $J$ is effectively changed by applying a pulsed linear
potential such that $F \tau=\pi$. 
Alternatively, one can understand the change of sign by looking at the effect of
the linear potential on the Hamiltonian in the Heisenberg picture. Indeed,
one can see that $a_j \rightarrow e^{i F \tau} a_j$. Thus,  
the interaction term remains unchanged, and the hopping term 
$a_i^\dagger a_{i+1} \rightarrow e^{i F \tau} a_i^\dagger a_{i+1} 
|_{\tau=\pi/F} = -a_i^\dagger a_{i+1}$.
In other terms, the linear potential creates a phase imprinting on the lattice
such that neighboring sites have a $F \tau$ phase difference.
Such a phase imprinting has already been demonstrated using a 
rapid displacement of the quadratic trapping potential \cite{Greiner}.
In general, a pulsed laser masked with a linear gradient \cite{PhillipsPhase} can be used. 
The linear phase imprint can be done either in one, two, or
three dimensions.

The complete sequence for the Loschmidt echo in the optical lattice is 
schematized in the inset of Fig. 1. 
The magnetic field ramp and the $\pi$ phase imprinting need to be done much 
faster than the characteristic time of the dynamics of the BHM:
typically, $\hbar/J \sim 1$ ms and $\hbar/U \sim 100$ $\mu$s \cite{Greiner}.
Notice that it is not necessary to measure the full overlap between
unknown states. Instead, only the probability of finding the system in the
initial state is needed.
I will discuss this issue in detail below,  after considering the effects of 
experimental errors in the LE.  

I will classify the sources of decay of the LE in three categories:
A first category consists of what I call {\em natural sources}:
external degrees of freedom that are not described by the simulated
Hamiltonian.
In this category I include e.g. coupling to environments, 
or truncated terms of the real Hamiltonian:
second neighbor hopping,
excitations between different bands, and so on. 
Natural decay sources give a bound on the fidelity of
the quantum simulation. 
A second category consists of {\em artificial sources}: external
fields or potentials purposely placed in the experiment. 
Since these are not (in principle) reversed by the LE, one could measure their 
strength through the decay of the echo. Below I will give an example of how to 
measure the gravity potential with the optical lattice. The third category 
consists of {\em experimental errors} in the implementation of the LE,
which limits the observation of decay due to natural and artificial sources.

The precision of $U \rightarrow -U$
is limited mostly
by the homogeneity of the magnetic field in the trapping region. 
Variances across the sample on the order of $10$ $mG$ 
or less are possible \cite{Jin04}.
For comparison, Feshbach resonances can be found at fields 
$B_0 \sim 10G-1000G$, with widths ${\cal O}(\Delta B) \simeq 1 G$. 
It might be useful to work near a broad Feshbach 
resonance in a region with low sensitivity to $B$. 
Assuming a homogeneous error $\delta U$, 
and an initial Fock state, for short times $t \ll \hbar/\sqrt{J \delta U}$ 
a perturbative expansion of the fidelity
$f(t)$, Eq. (\ref{loschmidt}),  gives $f(t)\simeq 1-J^2 \delta U^2 t^4$.
The fourth power contrasts with the typical quadratic decay
of perturbation theory \cite{Peres}. It appears because 
the interaction term of the BHM is diagonal in the Fock basis, 
and the second order terms cancel out. 
This expression is valid up to times $t \propto \hbar/J$, at which the second
order terms become relevant and a decay 
$f(t) \simeq \exp(-\beta \delta U^2 t^2)$ is observed numerically with $\beta={\cal O}(U/J)$ ,
see Fig. 1.

Perturbation theory predicts that errors in the $\pi$ phase gate, 
represented in the BHM by $J \rightarrow -J+\delta J$, cause a decay
$f(t)\simeq 1-\delta J^2 t^2$ for times $t \ll \hbar/\delta J$ (see Fig. 1).
Experimentally, the accuracy of the linear phase imprinting is limited
by fluctuations in the laser intensities, 
the precision of the imprinting mask and the lenses used to scale it
down to the size of the atomic cloud. Phase imprinting resolution of up to $\sim 5 \mu m$ 
has been achieved  \cite{PhillipsPhase}
(compared to lattice spacings $\sim 0.5 \mu m$), 
but this is for a sharp phase step. 
Linear smooth gradients and lattices with controllable intersite spacing \cite{Raizen}
can make the phase imprinting much more precise.  
Depending on the strength of natural or artificial decay sources of the LE, 
precision up to a few percent of $J$ could still be tolerable (Fig. 1). 
The duration $\tau$ of the phase imprinting pulse has to be short
enough so that there is no dynamics
inside each site of the lattice. 
By expanding a Wannier function of width  $~d/\pi$ 
in the eigenstates of the lineal potential, this condition can be cast as
$\tau < 2 m d^2/\pi^2 \hbar$. For typical values
$\tau < 1 \mu s$, within experimental reach \cite{PhillipsPhase}.

\begin{figure}[tb]
\centering \leavevmode
\epsfxsize 3.2in
\epsfbox{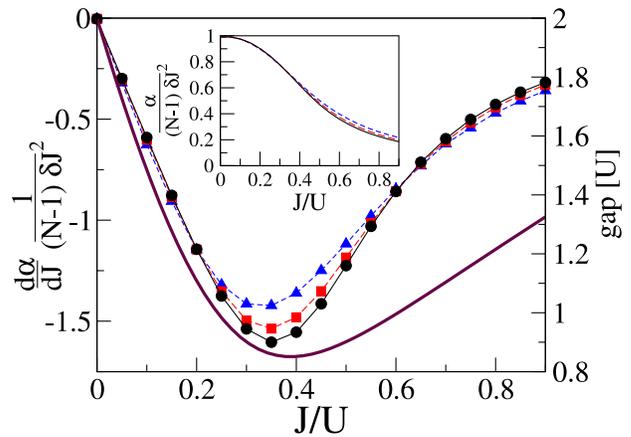}
\caption{(left vertical axis) Derivative of the initial decay rate $\alpha$ of the LE  
for two  Hamiltonians with $J-\delta J/2$ and $J+\delta J/2$,
as a function of $J/U$ and fixed $\delta J=0.05U$.
The plots are for $1$-D systems with $N=4$, $6$ and $8$ sites 
(triangles, squares and circles
respectively) and one particle per site. 
The curves are normalized to the $J=0$ value, $\left. \alpha\right|_{J=0}=(N-1)\delta J^2$. 
The rates are obtained by a Gaussian fit to $f(t)=\exp (-\alpha t^2)$.
A singularity in the derivative of $\alpha$  develops near the critical point of the transition.
For comparison, the gap between the ground and the first excited state for $N=8$ is shown
(thick line, right axis). In the thermodynamic limit, the transition point moves to $J=0.52U$
\cite{DMRG}. In the inset, the normalized decay rates $\alpha$ for the same parameters as the
main plot.} 
\label{Figure2}
\end{figure}

Perhaps the most challenging step of the LE experiment proposed here
is the preparation/detection of a particular many-body state. 
Some states, like the ground state of the superfluid phase, or selected Fock states,
can be faithfully prepared experimentally  \cite{Greiner,NISTMott}.
However, measuring the probability of finding the system in one of these 
particular states might not be simple. 
This could be done for Fock states 
if the occupancy of each site of the lattice can be individually measured:
First, the hopping dynamics is quenched with a sudden increase in the 
optical lattice potential. Then, the number expectation in each site is 
measured, collapsing the state in the Fock basis. The process is done many times,  
with the probability of finding each Fock state being its relative frequency. 
However, spatial periods of optical lattices are shorter than optical resolutions, thus
individual site addressability is still far from experimental reach. 
Clever techniques \cite{DasSarma} or
long wavelength lattices could be a solution to this problem.
Also, single site addressability might not be strictly necessary. 
Other schemes with indirect 
approaches could at least give bounds to the LE.
For instance, using microwave spectroscopy and 
density dependent transition frequency shifts,
the group of Ketterle recently imaged sites in the lattice with different
occupation numbers \cite{KetterleMott}. 
Instead of measuring the full LE, Eq.~(\ref{loschmidt}), one could prepare and measure
a state in a subspace of the full Hilbert space, i.e. by removing all
sites with occupancy 2. The resulting measurement gives a bound on the LE
determined by the relative size of the subspace considered.

If the BHM is faithfully simulated, the LE can have sensing applications.
For example, one can measure the strength of external potentials
({\em artificial} decay sources) whose interaction with the atoms
does not change sign with the LE procedure. 
The decay rate of the LE measures the strength of the potential \cite{Peres}.
Consider a vertical one dimensional optical lattice,
where the atoms are affected by an extra linear term given by 
gravity $V = d m g \sum_j j n_j$, with $m$ the mass of the bosons 
and $g$ the gravity acceleration. Time dependent perturbation calculation gives
$f(t) \simeq 1 - (2g d m J)^2 t^4$, supported by numerical results with different
initial states (see Fig. 1). The fourth power decay means that 
performing accurate measurements of $g$ with the LE could be difficult.
However, the LE in the BHM might be more sensitive to other artificial potentials, 
and thus be useful for metrology. As a side note, the spectrum of the BHM
presents many avoided crossings as a
function of $g$, and a regime with a Wigner-Dyson statistics of energy spacings.
In general this is a signature of quantum chaos \cite{qchaos}. 
The LE has been intensely studied in this subject  \cite{Jalabert}
and could be a powerful tool to investigate this problem.

The LE in the optical lattice can also be used to measure the critical point
of the superfluid-insulator transition in the BHM. 
The strategy is to use the algorithm for detecting quantum critical points with 
a one qubit quantum computer proposed in 
Ref. \onlinecite{CucchiettiQPT}. However, 
the present approach is simpler to implement because it does away with
the need of the qubit: the LE is implemented directly on the system.
The measurement of the 
critical point of the transition would be as follows: 
First, prepare and evolve the ground state 
of the BHM Hamiltonian (this could be relaxed to other states) with a given 
set of parameters $U$ and $J$. Second, create a ``slightly'' imperfect time reversal 
by a known small amount $\delta$, e.g. in $J$. 
Finally, measure the echo as a function of $J$, keeping $\delta$ fixed. 
In Fig. 2 the decay rate of the LE for short times and its derivative are shown
as a function of $J/U$ for a fixed perturbation $\delta J$. 
By increasing the size of the system, the derivative of the short time decay rate develops
a singularity at the critical point,
consistent with the results found in Ref. \onlinecite{Rossini} for a different class of systems.
In the thermodynamic limit, the $1$-D 
BHM-superfluid phase is a critical phase (Kosterlitz-Thouless transition).
Therefore, the BHM is an interesting system to understand what more
information can be obtained
about quantum phase transitions by studying the LE \cite{Zanardi}. 

In summary, I described how to reverse the time evolution of the Bose-Hubbard
model that describes ultracold bosonic atoms in an optical lattice.
Although its realization presents some challenges, it is within
reach of current or near-future experimental technology. 
I showed that the time reversal (Loschmidt echo) has many interesting 
applications, such as measuring the fidelity of the quantum simulation of
the BHM, measuring the strength of external potentials,
and even finding the critical point of the BHM quantum phase transition. 
It will be interesting to see if the LE can be further developed as a
sensing or a quantum information tool. 

I wish to thank Eddy Timmermans for many fruitful conversations, and 
in particular for suggesting the phase imprinting procedure.
I also acknowledge Philippe Jacquod, 
Juan Pablo Paz, Mark Raizen, and Paolo Zanardi for valuable discussions, and Diego Dalvit 
and Bogdan Damski for carefully reading the manuscript.


\begin{thebibliography}{99} 

\bibitem{Feynman} R. Feynman, Int. J. Theo. Phys., {\bf 21}, 467 (1982).

\bibitem{perturbations} More generally, the perturbation could be different
in the second part of the evolution, ${\cal H}^{'} =-{\cal H}_{sim}+V^{'}$. 
However, the results
in this paper remain the same replacing $V$ with some form of average between $V$ and
$V^{'}$.

\bibitem{Peres} A. Peres, Phys. Rev. A \textbf{30}, 1610 (1984); A. Peres,
in \textit{Quantum Chaos, }edited by H. Cerdeira, R. Ramaswamy, M. C.
Gutzwiller and G. Casati, (World Scientific, 1991).

\bibitem{CiracBHM} D. Jaksch {\em et al}, Phys. Rev. Lett. {\bf 81}, 3108 (1998).

\bibitem{Greiner} M. Greiner {\em et al}, Nature {\bf 415}, 39 (2002).

\bibitem{Lewenstein} M. Lewenstein {\em et al}, cond-mat/0606771.

\bibitem{Sachdev} S. Sachdev, {\em Quantum Phase Transitions} (Cambridge University Press, Cambridge, 1999).

\bibitem{PhillipsPhase} L. Dobrek {\em et al}, Phys. Rev. A {\bf 60}, 
R3381 (1999); Denschlag {\em et al}, Science {\bf 287}, 7 (2000).

\bibitem{Feshbach} S. Inouye {\em et al}, Nature {\bf 392}, 151 (1998); 
J.L. Roberts {\em et al}, Phys. Rev. Lett. {\bf 81}, 5109 (1998);
E. Timmermans {\em et al}, Phys. Rep. {\bf 315}, 199 (1999).

\bibitem{Hahn} E. L. Hahn, Phys. Rev. \textbf{80}, 580 (1950); R. G.
Brewer and E. L. Hahn, Sci. Am. \textbf{251}, 50 (1984).

\bibitem{Horacio} H.M. Pastawski {\em et al}, 
Physica A {\bf 283}, 166 (2000).

\bibitem{Israel} M. F. Andersen, A. Kaplan, and N. Davidson, 
Phys. Rev. Lett. {\bf 90}, 023001 (2003).


\bibitem{Jalabert} R. A.~Jalabert and H. M.~Pastawski, Phys. Rev.
Lett. \textbf{86}, 2490 (2001).

\bibitem{CucchiettiPRL} F.M.~Cucchietti {\em et al}, 
Phys. Rev. Lett. \textbf{91}, 210403 (2003).

\bibitem{Quan} H.T. Quan {\em et al}, Phys. Rev. Lett. {\bf 96}, 140604 (2006).

\bibitem{CucchiettiQPT} F.M. Cucchietti, S. Fernandez-Vidal, and J. P. Paz,  quant-ph/0604136.

\bibitem{Rossini} D. Rossini {\em et al}, quant-ph/0605051.

\bibitem{Zanardi} M. Cozzini, P. Giorda, and P. Zanardi, quant-ph/0608059.

\bibitem{Nielsen} M. A. Nielsen and I. L.~ Chuang, \textit{Quantum computation
and quantum information} (Cambridge University Press, Cambridge, New York, 2000).

\bibitem{Fisher} M. P. Fisher {\em et al}, Phys. Rev. B {\bf 40}, 546 (1989).

\bibitem{Raizen} T.P. Meyrath {\em et al}, Phys. Rev. A {\bf 71}, 041604R (2005).

\bibitem{PrivCom}  M. Boshier, private communication.

\bibitem{CiracOL} D. Porras and J. I. Cirac, Phys. Rev. Lett. {\bf 93}, 263602 (2004).

\bibitem{Jin04} S. Inouye {\em et al}, Phys. Rev. Lett. {\bf 93}, 183201 (2004)

\bibitem{NISTMott}  S. Peil {\em at al}, Phys. Rev. A {\bf 67}, 051603(R) (2003).

\bibitem{DasSarma} C. Zhang, S.L. Rolston, and S. Das Sarma, quant-ph/0605245.

\bibitem{KetterleMott} G.K. Campbell {\em et al}, cond-mat/0606642.

\bibitem{qchaos} F. Haake, {\em Quantum Signatures of Chaos}, 
2nd Edition (Springer-Verlag, Heidelberg, 2001).

\bibitem{Tonks} B. Paredes {\em et al}, Nature {\bf 429}, 277 (2004).

\bibitem{DMRG} S. Rapsch, U. Schollw\"{o}k, and W. Zwerger. Europhys. Lett. {\bf 46} (5),
559 (1999).

\end{thebibliography}
\end{document}